\documentclass{pasa}%

\usepackage{graphicx}

\def\Curtin{$^{1}$}
\def\AASTRO3D{$^{2}$}
\def\UToronto{$^{3}$}
\def\ASU{$^{4}$}
\def\UWisc{$^{5}$}
\def\UW{$^{6}$}
\def\UWA{$^{7}$}
\def\USyd{$^{8}$}
\def\WSU{$^{9}$}
\def\CSIRO{$^{10}$}
\def\UMelb{$^{11}$}

\def\la{\ifmmode\stackrel{<}{_{\sim}}\else$\stackrel{<}{_{\sim}}$\fi}

\title[MWA Phase II]{The Phase II Murchison Widefield Array: Design Overview}

\author[Wayth et al.]{Randall B. Wayth\Curtin\thanks{Email: r.wayth@curtin.edu.au},
Steven~J.~Tingay\Curtin,
Cathryn~M.~Trott\Curtin$^,$\AASTRO3D,
David~Emrich\Curtin,
Melanie~Johnston-Hollitt\Curtin,
Ben~McKinley\Curtin$^,$\AASTRO3D,
B.~M.~Gaensler\UToronto$^,$\AASTRO3D,
A.~P.~Beardsley\ASU,
T.~Booler\Curtin,
B.~Crosse\Curtin,
T.~M.~O.~Franzen\Curtin,
L.~Horsley\Curtin,
D.~L.~Kaplan\UWisc, 
D.~Kenney\Curtin,
M.~F.~Morales\UW, 
D.~Pallot\UWA,
G.~Sleap\Curtin,
K.~Steele\Curtin,
M.~Walker\Curtin,
A.~Williams\Curtin,
C.~Wu\UWA,
Iver.~H.~Cairns\USyd, % has opted in
M.~D.~Filipovic\WSU, % has opted in
S. Johnston\CSIRO, % has opted in
T. Murphy\USyd , % has opt-ed in
P. Quinn\UWA, % has opt-ed in
L. Staveley-Smith\UWA$^,$\AASTRO3D, % has opt-ed in
R. Webster\UMelb$^,$\AASTRO3D ~and % has opted in
J.~S.~B.~Wyithe\UMelb$^,$\AASTRO3D % has opt-ed in
\\
\affil{\Curtin International Centre for Radio Astronomy Research (ICRAR), Curtin University, Bentley 6845 Australia}
\affil{\AASTRO3DARC Centre of Excellence for All Sky Astrophysics in 3 Dimensions (ASTRO 3D)}
\affil{\UToronto Dunlap Institute for Astronomy \& Astrophysics, University of Toronto, 50 St George St, Toronto, ON, M5S 3H4, Canada}
\affil{\ASU School of Earth and Space Exploration, Arizona State University, Tempe, AZ 85287, USA}
\affil{\UWisc Department of Physics, University of Wisconsin--Milwaukee, Milwaukee, WI 53201, USA}
\affil{\UW Department of Physics, University of Washington, Seattle, WA 98195, USA}
\affil{\UWA International Centre for Radio Astronomy Research (ICRAR), University of Western Australia, Crawley, WA 6009, Australia}
\affil{\USyd Sydney Institute for Astronomy (SIfA), School of Physics, The University of Sydney, NSW 2006, Australia}
\affil{\WSU Western Sydney University, Locked Bag 1797, Penrith South DC, NSW, 1797, Australia}
\affil{\CSIRO CSIRO Astronomy \& Space Science, Australia Telescope National Facility, P.O. Box 76, Epping, NSW 1710, Australia}
\affil{\UMelb School of Physics, The University of Melbourne, Parkville, VIC 3010, Australia}
}%

\jid{PASA}
\doi{10.1017/pas.\the\year.xxx}
\jyear{\the\year}

\usepackage{aas_macros}
\usepackage{hyperref} 
\hypersetup{colorlinks,citecolor=blue,linkcolor=blue,urlcolor=blue}

\hypersetup{draft}
\begin{document}

\begin{frontmatter}
\maketitle

\begin{abstract}
We describe the motivation and design details of the ``Phase II'' upgrade of the Murchison Widefield Array (MWA) radio telescope. The expansion doubles to 256 the number of antenna tiles deployed in the array.
The new antenna tiles enhance the capabilities of the MWA in several key science areas.
Seventy-two of the new tiles are deployed in a regular configuration near the existing MWA core. These new tiles enhance the surface brightness sensitivity of the MWA and will improve the ability of the MWA to estimate the slope of the Epoch of Reionisation power spectrum by a factor of $\sim$3.5.
The remaining 56 tiles are deployed on long baselines, doubling the maximum baseline of the array and improving the array $u,v$ coverage.
The improved imaging capabilities will provide an order of magnitude improvement in the noise floor of MWA continuum images.
The upgrade retains all of the features that have underpinned the MWA's success (large field-of-view, snapshot image quality, pointing agility) and boosts the scientific potential with enhanced imaging capabilities and by enabling new calibration strategies.
\end{abstract}

\begin{keywords}
instrumentation: interferometers -- techniques: interferometric -- radio continuum: general -- radio lines: general -- early universe
\end{keywords}
\end{frontmatter}

%%%%%%%%%%%%%%%%%%%%%%%%%
%%%%%%%%%%%%%%%%%%%%%%%%%
\section{INTRODUCTION }
\label{sec:intro}
Radio astronomy at low radio frequencies ($<$500 MHz) has undergone a major renaissance over the last decade, with the establishment and operation of a new generation of large-scale observational facilities. The renaissance has been enabled by advances in signal processing and computing and has been motivated by challenging problems in astrophysics and cosmology.

Perhaps the most visible science challenge at low radio frequencies is the detection of the redshifted \textsc{Hi} signal from the early Universe, the so-called Cosmic Dawn (CD) and Epoch of Reionisation (EoR) \citep[e.g.][]{2006PhR...433..181F,koopmans15}.
A number of responses to this challenge have emerged, in the form of new radio telescopes and the refurbishment of existing radio telescopes.
For example, the Giant Metrewave Radio Telescope (GMRT; \citealp{2014ASInC..13..441G}) has undergone a number of upgrades. New telescopes such as the Low Frequency ARray (LOFAR; \citealp{2013A&A...556A...2V}) and the Murchison Widefield Array (MWA; \citealp{2013PASA...30....7T}), while strongly motivated by the EoR power spectrum experiment, are general purpose instruments capable of a wide range of investigations.
Other new instruments such as the Precision Array for Probing the Epoch of Reionization (PAPER; \citealp{2010AJ....139.1468P}), the Hydrogen Epoch of Reionization Array (HERA; \citealp{2017PASP..129d5001D}), and the 21 Centimeter Array (21CMA; \citealp{Zheng16}) are entirely motivated by, and designed for, the EoR power spectrum experiment. 
Collectively, these telescopes are making steady progress towards an EoR power spectrum detection over a range of redshifts \citep[e.g.][]{2011MNRAS.413.1174P,2015ApJ...801...51J,2015ApJ...809...61A,2016ApJ...833..102B,2017ApJ...838...65P}.

The nature of these instruments also provides us with a number of new practical challenges. The large data challenge is one. Other challenges include imaging and calibration over wide fields of view, wide fractional bandwidths, and taking account of complex propagation effects (e.g. ionospheric and interplanetary scintillation).
%While the data from these new facilities are readily amenable to standard radio astronomy interferometric data processing, meeting these challenges at the level to detect the EoR requires deep efforts and the development of new algorithms (references).

For those instruments designed to be more general purpose than just an ``EoR machine'', these challenges are more than compensated for by the vast new areas of parameter space opened up by the unique characteristics of these instruments: 1) large fields of view on the sky; 2) sensitive, high fidelity and high resolution imaging at low frequencies; 3) flexibility in choice of output data products; 4) large fractional bandwidths at low frequencies. Already, LOFAR and the MWA have proven the value of this new parameter space, with important new fields of research strongly emerging well before all the technical challenges have been completely met.

%MWA (plasma tubes, GLEAM, IPS, polarisation, pulsars etc). LOFAR (astroparticles, thunder storms, pulsars etc).

The future Square Kilometre Array (SKA) will build from the current generation of low frequency instruments, in particular the SKA precursors and pathfinders MWA, HERA, and LOFAR. In the vein of the MWA and LOFAR, the low frequency SKA (SKA Low) will have a wide range of science goals and will open up an even wider parameter space for discovery.

The MWA is located at the same remote and radio quiet site as SKA Low will be, the Murchison Radio-astronomy Observatory (MRO).
The design, construction, commissioning, and operation of the MWA has generated deep experience that is directly relevant to SKA Low.
%, which has closely informed the SKA design and planning process (references).
In addition, the MWA has been directly utilised as a development platform to prototype and test various sub-systems for SKA Low \citep[e.g][]{2015ITAP...63.5433S,8104992,2017PASA...34...34W}.

%The MWA is well placed to take further steps toward the SKA, through upgrades and extensions. 
Low frequency radio telescopes such as the MWA are naturally suited to upgrades, since the front end infrastructure and antenna systems are relatively simple and the backend signal processing and data processing are driven by the ever increasing capabilities of the commodity electronics market. The design philosophy of the MWA, incorporating commercial off-the-shelf  components where possible \citep[e.g.][]{2015PASA...32....6O,2015PASA...32....5T}, has led to a flexible and intrinsically extensible instrument that is responsive to the dynamic scientific environment created when a new and large parameter space is entered.

Here we describe the next steps in the evolution of the MWA, with a doubling of the number of antennas in the array, a doubling of the maximum baseline in the array, and an exploration of calibration techniques based on alternative array architectures. These front-end upgrades address a number of different science areas, in keeping with a general purpose and flexible instrument, and pave the way for future upgrades that will be more focused on the signal and data processing in the back-end. The material described in this paper demonstrates that once the investment in the basic infrastructure has been made, substantial upgrades to the scientific capabilities of instruments like the MWA can be made in a highly efficient and cost-effective manner. We expect similar benefits will apply to the SKA Low telescope.

In this paper we refer to the original system that was deployed and commissioned in 2012/2013 as the ``Phase I'' MWA, which is described by \citet{2013PASA...30....7T}.
%Section \ref{sec:upgrade_details} describes the details of the upgrade, then Section \ref{sec:motivation} describes the main scientific motivations behind the upgrade.

%%%%%%%%%%%%%%%%%%%%%%%%%%%%%%%%%%%
%%%%%%%%%%%%%%%%%%%%%%%%%%%%%%%%%%%
\section{SCIENTIFIC MOTIVATION FOR THE UPGRADE}
\label{sec:motivation}

The Phase I MWA was motivated by a vast array of science goals spanning initial observations of the EoR power spectrum, Galactic and Extragalactic continuum and polarimetry, solar and ionospheric science and detection of transient sources \citep{2013PASA...30...31B}. In addition to covering the initial science goals, the Phase I MWA also produced a number of unforeseen results including the first detection of plasma tubes in the ionosphere \citep{Loi15a,Loi2015b}, the potential to ulitise the MWA for space debris detection and tracking \citep{Tingay2013b,Zhang2018}, tentative detections of new molecules \citep{2017MNRAS.471.4144T}, and use of interplanetary scintillation to identify and measure sub-arcsecond compact components in low-resolution, low-frequency radio surveys without the need for long-baseline interferometry or ionospheric calibration \citep{2015ApJ...809L..12K,Morgan2018, Chhetri2018}. The wide field of view and pointing agility of the MWA have also proven invaluable for both targetted and archival searches (from serendipitous observations) for radio emission associated with astronomical events such as gamma-ray bursts \citep{2015ApJ...814L..25K}, neutrino detections \citep{2016ApJ...820L..24C}, gravitational wave events \citep{2016ApJ...826L..13A,2017ApJ...848L..12A} and interstellar visitors \citep{2018ApJ...857...11T}. 

Given the successes of the Phase I MWA, when considering an upgrade to the facility, scientists generally desire improved capability in all key performance metrics of the telescope (sensitivity, angular resolution, frequency coverage, frequency resolution, time resolution, number of beams etc). Budgetary, timeline and practical constraints for the Phase II upgrade meant that the focus was put on expanding the physical array, with future upgrades to consider receivers and downstream digital systems.

The design of the Phase II array thus focuses on adding antenna tiles to improve capability for MWA Key Science, and to enhance capabilities in science areas where the MWA is a particularly well suited instrument.
Broadly, this comes down to 1) improving the sensitivity (and calibratability) for the EoR power spectrum measurement, and 2) improving the imaging properties of the array through improved angular resolution, $u,v$ coverage and reduced confusion.

\subsection{Epoch of Reionisation}

When the Phase I MWA was deployed, it was one of the first generation of telescopes aiming to detect the EoR power spectrum. It was designed as a general-purpose array with emphasis on imaging capability and brightness sensitivity on the angular scales relevant to the EoR. In the years following the initial design and deployment, it has become increasingly apparent that the dynamic range requirements (e.g. accuracy of calibration, accuracy of sky model etc) of the EoR detection experiment are just as challenging as the sensitivity requirements \citep{jacobs16, trott16, 2016ApJ...833..102B, barry16, 2016MNRAS.463.4317P, 2017ApJ...838...65P}.

In contrast to the MWA, the PAPER array \citep{parsons10} pursued a more targeted EoR power spectrum experiment, using a regular grid of simple dipole antennas. The regular antenna arrangement is especially useful for the EoR power spectrum, because the many identical (``redundant'') baselines formed by a regular grid 
boost the sensitivity for a power spectrum measurement \citep{2012ApJ...753...81P}.
A regular array configuration also enables one to exploit the benefits of redundant calibration. This removes some dependence on having an accurate sky model to calibrate the antenna gains. The under-construction HERA telescope \citep{2017PASP..129d5001D} also uses a regular configuration.

Motivated by this experience with existing EoR telescopes and experiments, 72 of the new antenna tiles were deployed in a regular hexagonal configuration. Section \ref{sec:upgrade_details} provides details of the deployment and section \ref{sec:capability} discusses the performance implications of the upgrade. 

%MWA Phase II was designed with the key choices for an interferometric low-frequency array to undertake EoR power spectrum science: (1) higher brightness sensitivity on arcminute-degree scales (dense packing of baselines of 10--100~m length), (2) high-resolution, deep sky model (compact and extended sources) for calibration and foreground signal subtraction, and (3) a stable and calibratable instrument.
%The MWA upgrade delivers these capabilities.

%The two principal features of the upgrade (new regular hex-configuration tiles and new long baselines tiles) are thus directly relevant to EoR power spectrum science.

%%%%%%%%%%%%%%%%%%%%%%%
\subsection{Image-based science}
Aside from raw sensitivity, the capability of a radio interferometer is usually constrained by its angular resolution and by the spatial scales that are measured given the antenna layout.
The Phase I MWA layout is a hybrid configuration having baselines ranging in length from approximately 7 to 2800 metres, with many tiles in the core region. The synthesised beam size of the Phase I array is approximately 2 arcmin at 150\,MHz.
Improving the angular resolution of the MWA for Phase II was identified for its potential to impact a broad range of science areas.
In addition to the ability to better resolve objects and structures within sources, improved angular resolution directly impacts the classical and sidelobe confusion in continuum images. 

For Phase II, 56 new antenna tiles were deployed beyond the circumference of the Phase I array. The new tiles double the maximum baseline of the array and generate more uniform $u,v$ coverage. The new tiles require additional hardware to transmit signals over long distances as described in Section \ref{sec:longbaselines}. The performance implications of the new tiles are discussed in section \ref{sec:capability_imaing}.

%%%%%%%%%%%%%%%%%%%%%%%%%%%
%\subsection{The bridge to SKA Low}

%%%%%%%%%%%%%%%%%%%%%%
%%%%%%%%%%%%%%%%%%%%%%
\section{THE PHASE II UPGRADE}
\label{sec:upgrade_details}
%- constraints and goals of the upgrade (existing tiles, infrastructure, land boundaries, correlator etc)\\
%- summary of main features \\

The existing MWA consists of 16 in-field receiver units, each of which services eight antenna tiles \citep[see][]{2013PASA...30....7T}.
The Phase II upgrade adds an additional 128 tiles, taking the total number of deployed tiles to 256.
The 128 new tiles comprise:
\begin{itemize}
\item 72 tiles arranged in two regular hexagonal configurations near the existing MWA core (Fig. \ref{fig:arr_compact}); and
\item 56 long baseline tiles placed beyond the limit of the existing array (Fig. \ref{fig:arr_extended}), taking the maximum baseline for the Phase II array to approximately 5.3\,km.
\end{itemize}

The existing receivers \citep{2015ExA....39...73P} and correlator \citep{2015PASA...32....6O} were retained for this expansion, which limits the total number of tiles that can operate at any time to 128\footnote{No new receiver hardware is available for Phase II, however potential receiver upgrades are under development as noted in section \ref{sec:conc}.}. As such, the Phase II MWA will be periodically reconfigured between \emph{compact} and \emph{extended} configurations, each of which contains 128 operating tiles. Reconfiguration involves physically disconnecting tiles from some receivers, relocating the receivers, then connecting new tiles\footnote{The reconfiguration takes approximately 2 weeks and happened approximately once per year between mid 2016 and mid 2018. Reconfigurations are anticipated to be twice yearly moving forward}.

%David's suggested re-order text:
The overall goal was to have the maximum  baseline of  the  Phase II array be as large as possible while retaining the  excellent snapshot $u,v$ coverage that has underpinned much  of the MWA's success. Placement of the long baseline  tiles was constrained by a number of practical and logistical considerations.

A feature of the original MWA deployment that has proven invaluable for the Phase II expansion is the excess power and communications infrastructure that was initially installed. The MWA's in-field receivers are placed on concrete pads next to service points for buried power and communication infrastructure. During the infrastructure work for the initial MWA deployment it was decided to install extra buried infrastructure beyond what was needed for Phase I. The reason was that the additional cost of service infrastructure was small compared to the overall infrastructure costs, and this excess capacity would be useful for future expansion. During Phase II construction extra concrete pads were placed near the central region of the MWA, thereby creating sufficient space to utilize the extra buried capacity to host all 16 receivers near the core, which is required for the new compact configuration.

When choosing which existing tiles were to be used in either of the new configurations, we opted to include or exclude all eight tiles on a receiver (hence, we retain or change all tiles connected to a receiver) for practical reasons. Hence, the two new configurations consist of a subset of the receivers from the Phase I array (and all their associated tiles), plus new tiles, which are connected to relocated receivers.

\subsection{Compact hexagonal configuration antenna tiles}
\label{sec:hexes}
\begin{figure}
\vspace{-1cm}
\includegraphics[width=\linewidth, height=7.8cm]{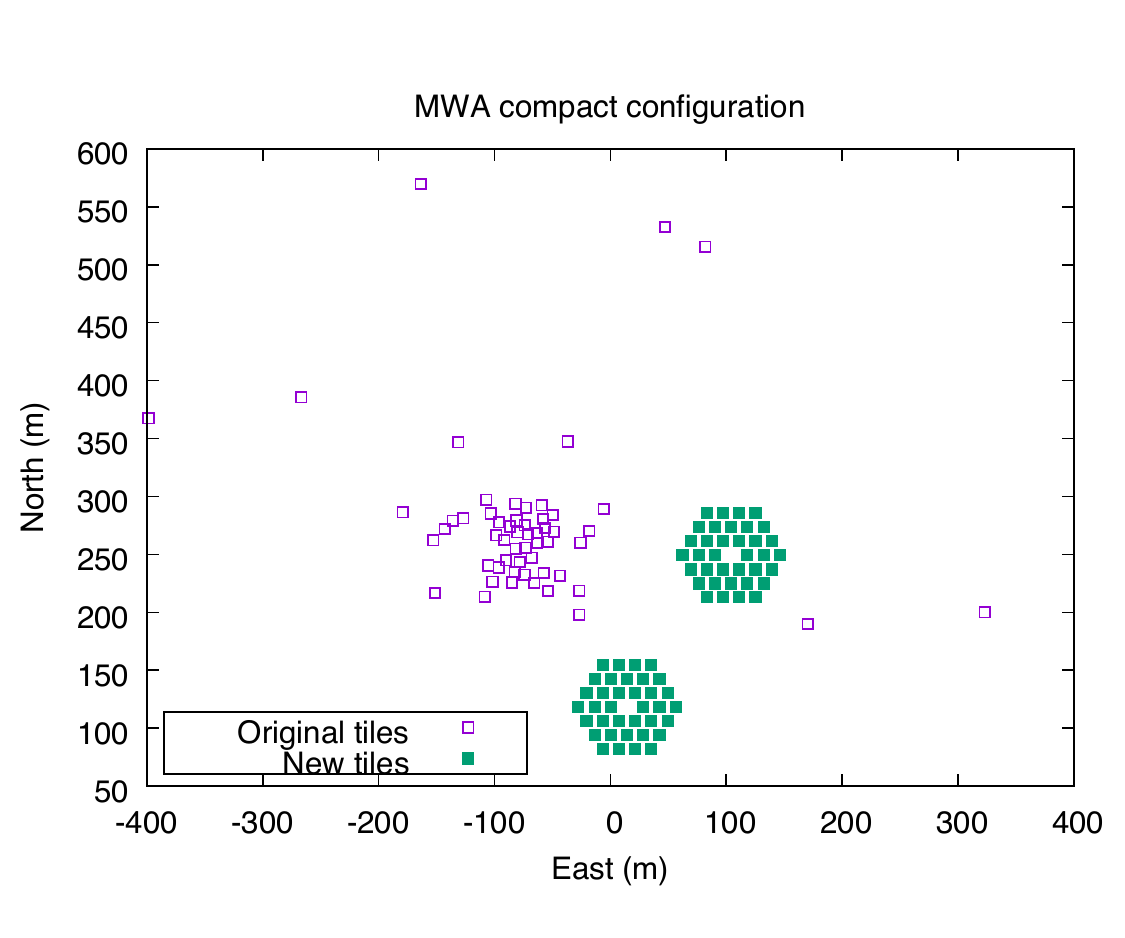}
\vspace{-0.8cm}
\caption{The compact configuration including the 72 new tiles arranged as two regular hexagonal arrays (filled squares). Note the size of the squares is not to scale.}
\label{fig:arr_compact}
\end{figure}

The compact configuration consists of seven Phase I receivers and their corresponding 56 tiles, which comprise the bulk of the existing MWA core tiles, plus the 72 new hexagonal configuration tiles. The nine receivers required for the new tiles are relocated from elsewhere in the array, mostly from the existing Phase I outermost tiles.
The layout of the compact configuration is shown in Fig~\ref{fig:arr_compact}.

The vast majority of the baselines in the compact configuration are shorter than 200\,m, consistent with requirements for an EoR power spectrum detection being the primary science driver for this configuration. Clearly the compact configuration is a hybrid between the regular hexagonal configuration tiles and the pseudo-random core of the existing array, with the regular hexagonal configurations providing many redundant baselines. 

The hexagonal configuration of new tiles was chosen over other possibilities (such as a rectangular grid, which would have better theoretical sensitivity for the power spectrum) for several reasons. Firstly, the hexagonal configuration has less pronounced grating features compared to a rectangular grid; the configuration matches the HERA configuration, hence will allow comparison of results for the same modes; the size of the hexagons is a good match to the size of the existing Phase I core region; finally, the hexagonal arrangement avoids potential issues with a rectangular grid's grating features beating against the primary beam of the MWA's square tiles.

\subsection{Long baseline tiles}
\label{sec:longbaselines}
\begin{figure}
\vspace{-0.6cm}
\includegraphics[width=\linewidth]{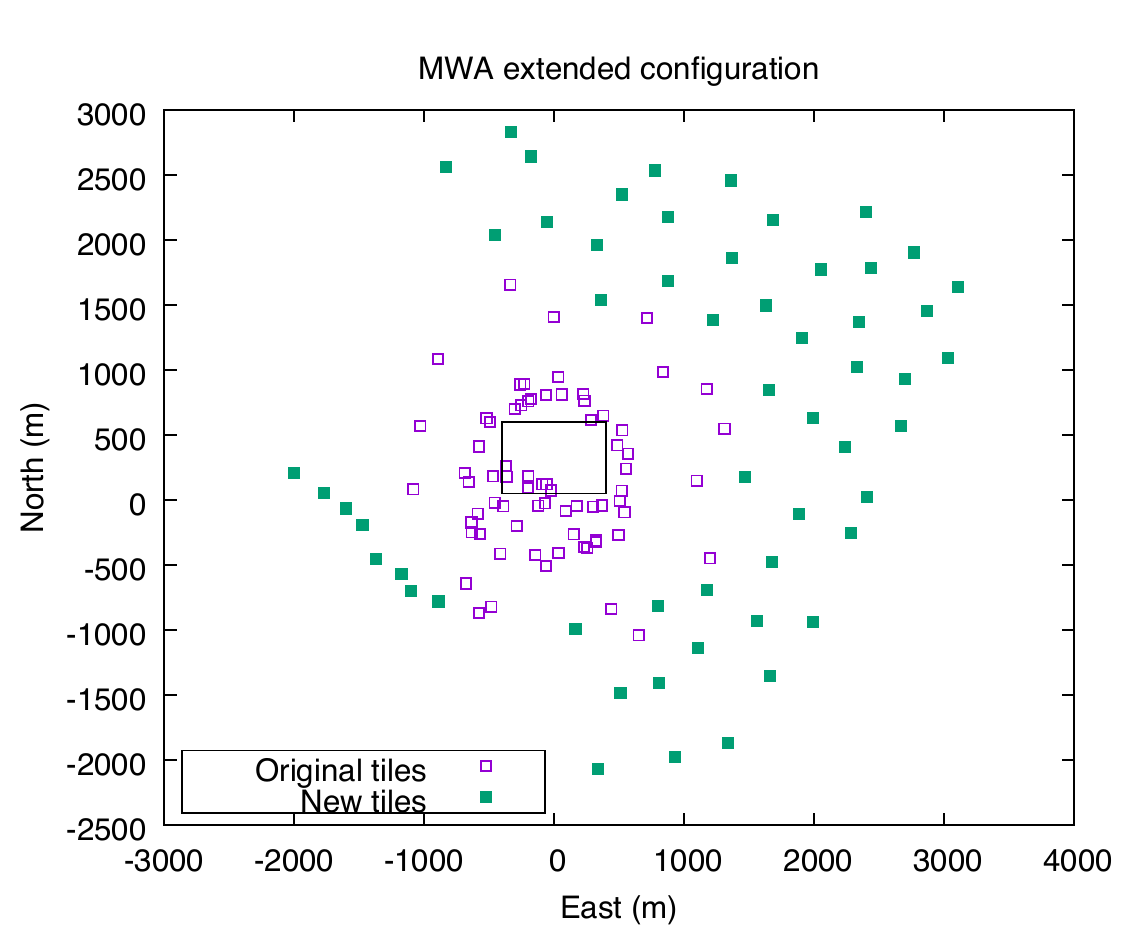}
\caption{The extended configuration including the 56 new long baseline tiles (filled squares). The inset rectangle shows the area bounded by Fig. \ref{fig:arr_compact}.}
\vspace{-0.38cm}
\label{fig:arr_extended}
\end{figure}

The extended configuration consists of nine existing receivers (essentially the Phase I array minus the core) and their corresponding 72 tiles, plus the 56 new long baseline tiles.
The layout of the extended configuration is shown in Fig \ref{fig:arr_extended}.
The seven receivers required to service the new long baseline tiles are relocated from the core region.

All existing MWA tiles (and the new hex configuration tiles) are serviced by twin coaxial cables, which carry power and communications to the MWA beamformers, and the radio-frequency (RF) signal back from the tiles. The new long baseline tiles are substantially further from the nearest receiver pad than can be feasibly reached via coaxial cable, hence the new long baseline tiles use self-contained solar power units and RF-over-fibre technology to transmit the RF signal back to receivers.

Significant drivers for the cost and schedule of the expansion project were environmental and local heritage clearances, which are required if the land is disturbed by clearing and/or digging. As such, the additional infrastructure work to support the long baselines was limited to extending the existing access tracks; no new buried cables were installed. All new equipment for the long baseline tiles (including the tiles themselves, power supplies and fibre cables) was placed on the surface.

The locations of the long baseline tiles were determined by initially selecting target locations within existing boundaries of the site, such that the baselines were as long as possible and the resulting $u,v$ coverage was sensible. Each target location was visited and inspected such that a suitably clear, flat site near the target location could be chosen.
For all but two of the long baseline tiles, a suitable site was found within approximately 50 metres of the original target location, the remaining two were moved by several hundred metres to avoid locally sensitive areas.
This process of on-the-ground micro-siting was critical in meeting the cost, schedule and compliance requirements of the upgrade project.

\subsection{Power and signal transport for long baseline tiles}
\begin{figure*}
\includegraphics[width=\linewidth]{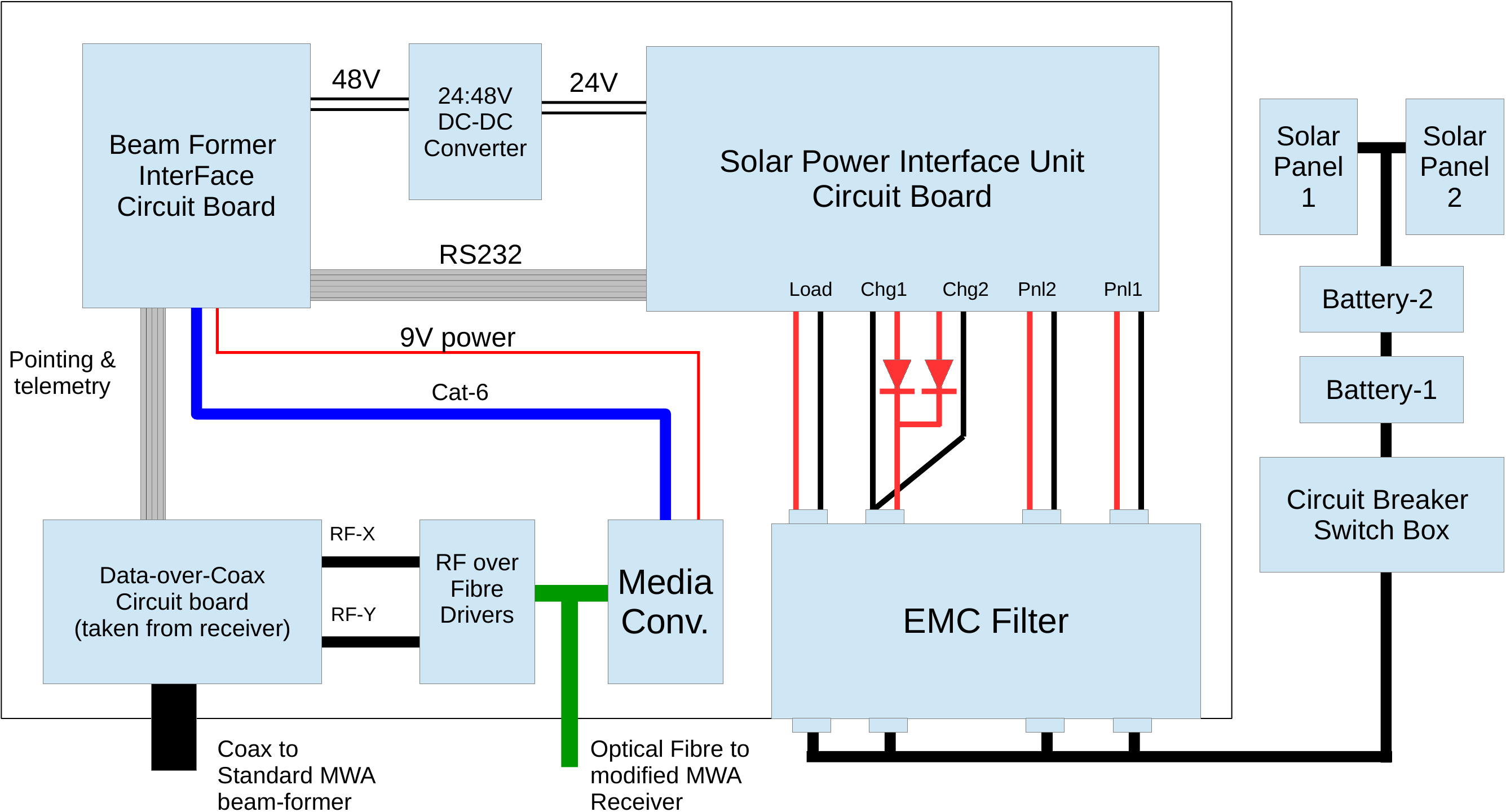}
\caption{Block diagram of long baseline tile power, control and signal transport system.}
\label{fig:BFIFetc}
\end{figure*}

All Phase I tiles (and the new Phase II hex configuration tiles) have beamformers that connect to MWA receivers via two coaxial cables that range in length up to a few hundred metres. The cables supply power to beamformers, carry RF signals from the beamformers, and provide two-way communications via a simple serial protocol. This system is supported by custom-designed ``data over coax'' (DoC) circuit boards, one of which is in the beamformer (the beamformer-side DoC), and the other is in the receiver (the receiver-side DoC).

The long baseline tiles use standard MWA beamformers. In order to operate these beamformers, supporting infrastructure that mimics the functionality provided by the receiver was required. A bespoke system was designed to do this, as illustrated in Fig. \ref{fig:BFIFetc}.

The long baseline tiles are connected to the rest of the MWA via a 6-core optical fibre cable. This cable provides standard two-way ethernet communications (using two fibres) and carries the RF signals from the tiles (on another two fibres, hence there are two spares). Power and communications with the tiles are provided via the Beamformer Interface (BFIF) unit, which includes a Raspberry Pi single board computer (SBC). The SBC controls a receiver-side DoC board, which in turn connects to the beamformer via short lengths of coaxial cable. The dual polarisation RF output from the DoC board is fed into two RF-over-fibre drivers (ASTRON model 350).

Power is provided by two 315\,W, 24\,V solar panels (Suntech STP315/24/VEM), which are mounted on a fixed frame nearby the tile. Four 150\,Ah, 12\,V lead-acid batteries provide energy storage, arranged as two sets of 24\,V batteries. The power subsystem is controlled by a solar power interface unit, which provides a low level monitor and control serial interface that is connected to the SBC. 

A key requirement for the power subsystem is to be radio quiet, so a typical off-the-shelf solution (which commonly uses rapid switching) was not feasible. Instead, a  custom low-power low-frequency switching supply was designed to trickle charge the batteries while also powering the tile. The trickle charge regulators can be bypassed under software control, allowing one or both panels to be directly connected to the batteries to provide a more rapid charge, depending on battery charge state and solar panel output. Software on the SBC monitors and controls this subsystem closely to ensure optimum battery lifetime, and additionally will shut down most of the loads if required to protect batteries from over-discharge. Finally, a carefully selected and tested off-the-shelf medium-frequency switch-mode power supply is used internally to convert the 24\,V from the panel/battery sub-system to the 48\,V required for the beamformer.

\begin{figure}
\includegraphics[width=\linewidth]{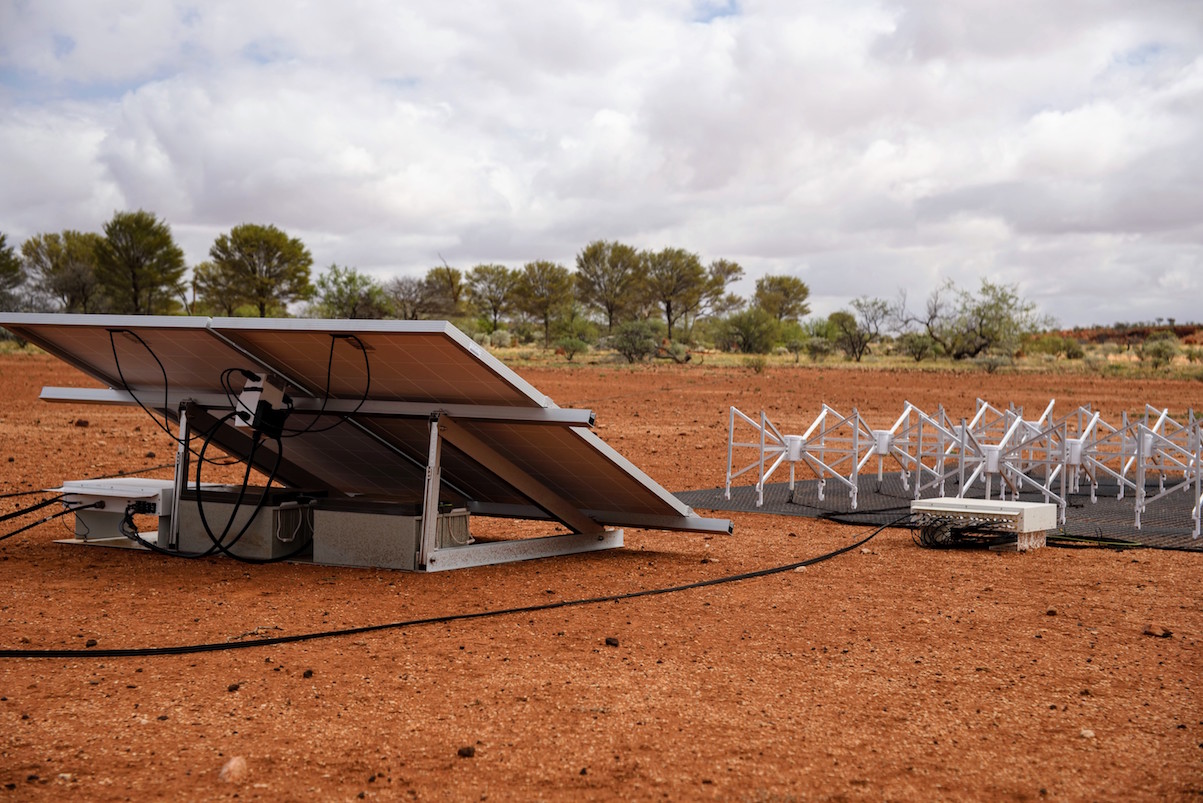}
\caption{A Phase II long baseline tile. The batteries and beamformer interface systems are located under the solar panels.}
\label{fig:PhII_Tile}
\end{figure}

Given the proximity of this solar power subsystem to its nearby antenna tile, we undertook careful electromagnetic compatibility (EMC) design to ensure that any radio emission generated by the charging electronics would not interfere with the astronomical signals. We therefore designed the enclosure to be very similar to the beamformer enclosure, which is known to provide good shielding from radiated emissions, and built a custom EMC filter to prevent any emissions from leaving the enclosure on power cables. Figure \ref{fig:PhII_Tile} shows an image of a Phase II long baseline tile.

Modifications were also required to the receivers connected to long baseline tiles.
%The seven sets of eight long baseline tiles are connected to seven modified MWA receivers, each receiver handling eight tiles.
An 8-port network switch with optical ports was fitted in each modified receiver to distribute the control and telemetry signals for the tiles. Also, the original eight receiver-side DoC boards were removed and replaced with a 16-channel RFoF receiver board, which converts the 16 optical RF signals back to electrical signals. These were then connected to the existing Analogue Signal Conditioner (ASC) modules. The remaining components of the receivers were unchanged.

\subsection{Correlator and archive}
\label{sec:correlator}

The digital systems of the MWA remain the same for this upgrade, although the new longer baselines do have an effect on the operation of the correlator and on-site archive.
Phase I observations were typically performed with time and frequency averaging in the correlator set to keep the output data rate within the capabilities of the archive. For example, observations requiring 0.5\,s time resolution were required to  average the native 10\,kHz frequency resolution down to 40\,kHz. Likewise, observations requiring the full 10\,kHz frequency resolution were required to time-average the data to 2\,s.

The Phase I correlator \citep{2015PASA...32....6O} does not fringe track. While this greatly simplifies the requirements of the correlator itself, the natural fringe frequency of the longest baselines sets an upper limit to the time averaging that can be performed in the correlator. Similarly, cable length differences that are uncompensated before the correlator set an upper limit to the frequency averaging that can be performed in the correlator.
The net effect of this is that for the extended array, the correlator must be run with higher time and frequency resolution as compared to Phase I.
Typically this is 0.5\,s time resolution and 10\,kHz frequency resolution. Compared to Phase I, this increases the maximum data rate generated by the correlator by a factor of 4.

To handle the larger data rate, two enhancements were made to the correlator and archive: expansion of the on-site archive system, and enabling in-situ compression of visibility data in the correlator.
The onsite archive \citep{2013ExA....36..679W} was upgraded by replacing the two existing data servers with six Dell PowerEdge R730xd servers (each with approximately 66\,TB of usable storage), and upgrading the switching network between the correlator machines and archive servers.

The in-situ compression of data takes advantage of the fact that the raw correlation coefficients generated by the correlator are integers, since the data going into the correlator are 4-bit integers \citep{2015PASA...32....6O}. Although the xGPU library performs the cross-multiply and accumulation as floating point numbers, the resulting correlation coefficients are simply integers with very small floating-point accumulation/rounding errors. Thus, instead of storing the coefficients as floating-point numbers, they are rounded and stored as 32-bit integers using the in-built Rice compression functionality provided by the CFITSIO library$\footnote{\url{https://heasarc.gsfc.nasa.gov/fitsio/fitsio.html}}$. This transformation is lossless. For typical observations, the correlation coefficients are less than 10000 or so, and hence the 32-bit integers are highly compressible, whereas the same number stored in floating point format would not be. For typical observations, this scheme reduces the data rate out of the correlator by approximately 2.5:1 compared to uncompressed data.

The downstream post-processing tool \textsc{Cotter} \citep{2015PASA...32....8O} (which applies delay corrections and creates a phase centre for the data, in addition to flagging) provides functionality for time and/or frequency averaging such that end users may not have to deal with the larger data volumes when the data are being converted from raw format (stored in FITS files) to a standard radio astronomy format. In addition, the compression feature specified in the FITS standard is handled transparently by all tools; hence no modifications to downstream tools were required.

%%%%%%%%%%%%%%%%%%%%%%%
%%%%%%%%%%%%%%%%%%%%%%%
\section{Capability of the upgraded MWA}
\label{sec:capability}

%%%%%%%%%%%%%%%%%%%%%%%%
\subsection{Calibration using redundant baselines}
\label{sec:EoR}
%The bandwidth of the MWA allows exploration of the EoR principally in the redshift range, $z=6-10$, encompassing a period thought to be dominated by a neutral hydrogen emission signal on scales of arcminutes--degrees \citep{koopmans15,mellema13}. At these redshifts, the signal is weak (10s~mK) compared with astrophysical foregrounds, but the sky temperature is lower than at higher redshifts. Despite its weakness, and in the absence of systematic errors and contamination, the signal (spatial fluctuations of the 21~cm brightness temperature) is theoretically detectable with 100s-1000s hours of data. However, the experiments are constrained by our imperfect knowledge of the sky and instrument, making data calibration and foreground signal subtraction extremely important for EoR science \citep{jacobs16,trott16,beardsley16,barry16,beardsley13}.

%Both LOFAR and MWA have demonstrated the importance of extremely high-fidelity sky models for calibration and foreground subtraction, and a smooth and well-understood system spectral response 

Redundant calibration, whereby the observation of the sky using many identical baselines recovers the antenna gains without requiring a sky model \citep{wieringa92,liu10,zheng14,dillon16}, was used by the PAPER telescope and is central to the proposed design for HERA.
Despite its promise of more accurate calibration by removing the dependence on a sky model, redundant calibration has its own set of challenges for real instruments with differing antenna responses, and suffers from degeneracies in some situations.
The MWA Phase II compact configuration addresses these challenges by combining information from redundant and non-redundant baselines, allowing a more holistic approach to calibration while also providing information about the variations between different antenna tiles \citep{2018ApJ...863..170L}.
%The other advantage of the redundant hexagonal subarrays is their high surface brightness response to EoR scales. Concentrating data collection on these scales yields high sensitivity and a consequent reduced total observing time.

%\cmt{perhaps we can cite Wenyang here, and Ronniy if his paper is submitted. At least we might be able to reference Ronniy's IAU conference proceedings.}

\subsection{Power spectrum sensitivity}
\begin{figure}
\includegraphics[width=\linewidth]{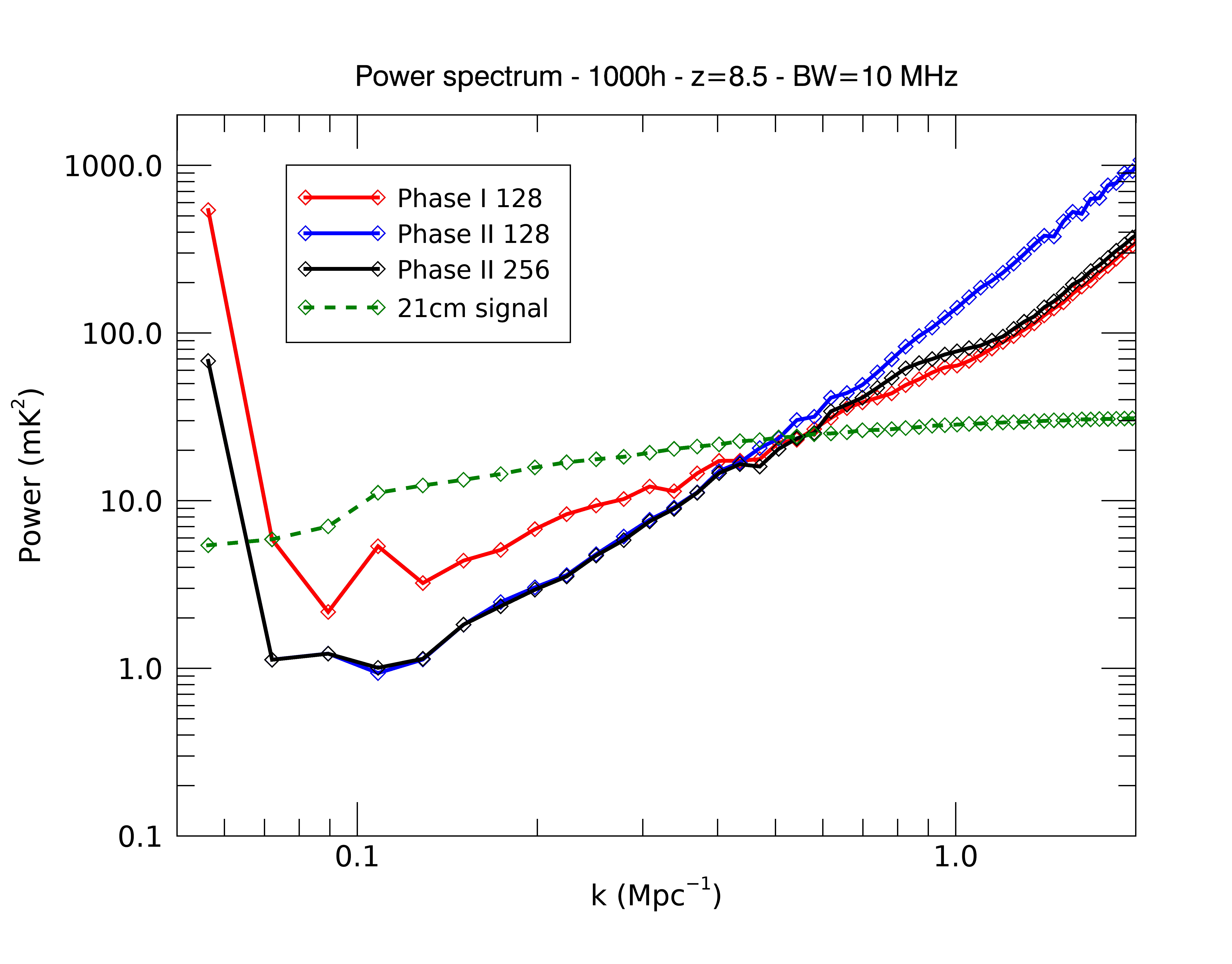}
\caption{Typical EoR power spectrum model at 150\,MHz with associated noise levels available to the Phase I and Phase II arrays with 1000\,h observation. ``Phase II 256'' represents the result from a future MWA upgrade where all 256 tiles are used simultaneously.}
\label{fig:eor_snr}
\end{figure}

Figure \ref{fig:eor_snr} shows the theoretical noise levels for a 1000~hour observation with the original MWA and the upgraded MWA, with the extra sensitivity at small $k$-modes, for a typical 21~cm signal model at 150\,MHz \citep[using 21cmFAST;][]{mesinger11}.
Over the range of scales of relevance to the EoR power spectrum at $z\sim 8$ ($u=[2,200]$), the raw sensitivity increase is a factor of four, as can be observed in the figure.

To provide a realistic assessment of the performance improvement for Phase II, a realistic foreground model is included in the analysis. For both Phase I and II, compact sources are assumed to be subtracted to five times the classical confusion level, with the lower confusion level for Phase II accounted for. Phase II therefore offers both improved sensitivity on large scales and lower foreground contamination. (Figure \ref{fig:eor_snr} includes a future ``Phase II 256'' array where all 256 antenna tiles are correlated.)
Note that Phase II arrays with both the compact configuration of 128 tiles and a full extended 256-tile array yield comparable performance on the large angular scales of importance for EoR science. Finally, we perform a Fisher Analysis of the theoretical signal-to-noise ratio attainable on the slope and offset of the power spectrum (in the presence of realistic foregrounds), yielding an expected improvement in parameter estimation performance by a factor of 3.5 for Phase II compared with Phase I.

%%%%%%%%%%%%%%%%%%%%%%%%%
%%%%%%%%%%%%%%%%%%%%%%%%%
%\section{DISCUSSION}
%\label{Discussion}
%- deployment issues and accuracy \\
%- expected performance of hexes for PS science \\
%- expected performance of long baselines for continuum science. Sensitivity, PSF size, quality \\
\subsection{Imaging}
\label{sec:capability_imaing}
\begin{figure*}
\includegraphics[width=8.5cm]{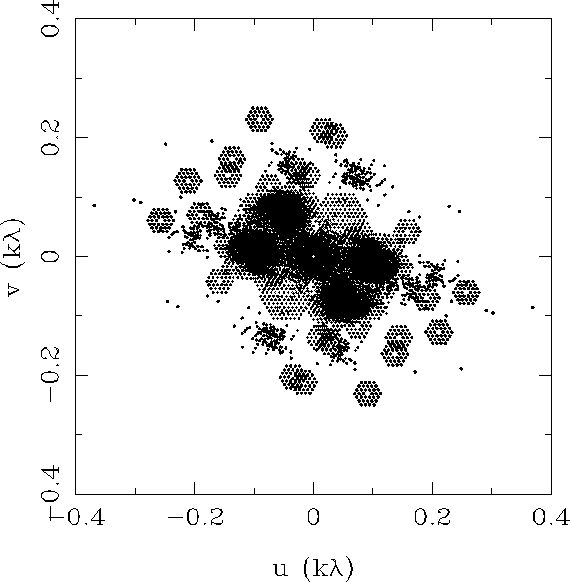}
\includegraphics[width=8.3cm]{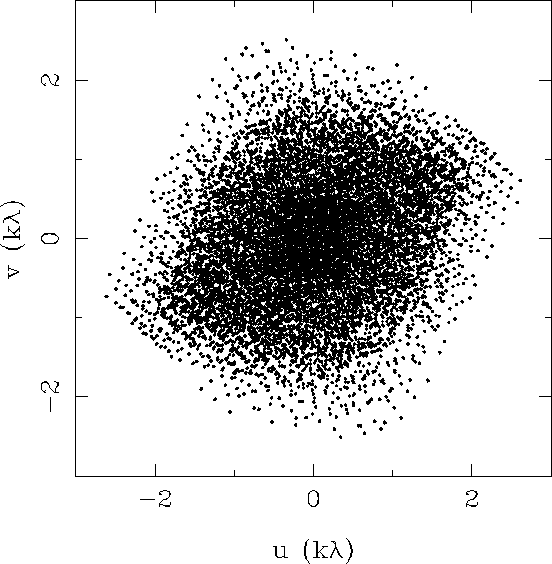}
\caption{Snapshot monochromatic $u,v$ coverage of the Phase II configurations at 154\,MHz. The left panel shows the compact array (zenith pointed) and the right panel the extended array (zenith pointed).}
\label{fig:uv}
\end{figure*}

Since both the compact and extended arrays contain 128 tiles, the theoretical radiometric sensitivity of MWA Phase II is the same as the Phase I array. In practice, however, the spatial resolution and $u,v$ coverage of the Phase II arrays mean the two configurations perform very differently depending on the application. The snapshot monochromatic $u,v$ coverage of the compact and extended configurations are shown in Figure \ref{fig:uv}.

\begin{figure}
\includegraphics[width=\linewidth]{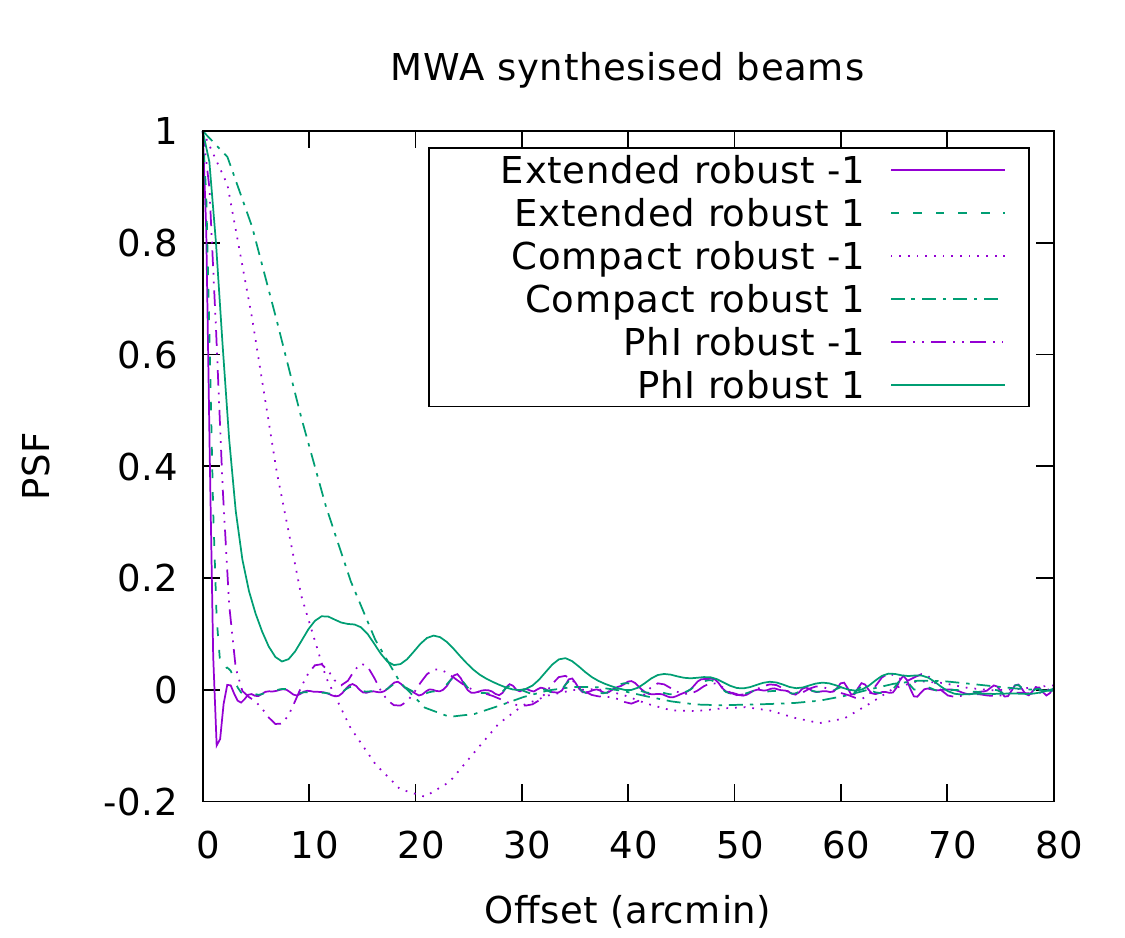}
\caption{Example cuts of the synthesised beam (zenith pointed) at 154\,MHz for the Phase I and Phase II arrays with different robust weights. In this example 1\,MHz of bandwidth is used.}
\label{fig:beamcuts}
\end{figure}

The extended configuration doubles the maximum baseline length compared to the Phase I array, and also does not contain the many short baselines in the core. This means the $u,v$ plane is more uniformly filled compared to the Phase I array, with an associated improvement in the synthesised beam.
Fig. \ref{fig:beamcuts} shows example 1-D cuts through the MWA's synthesised beam for the compact and extended configurations and the Phase I array for two different visibility weighting schemes, roughly corresponding to uniform weights (robust -1) and natural weights (robust 1). The example uses only 1\,MHz of bandwidth, so corresponds closely to the ideal monochromatic synthesised beam.

For continuum imaging compact sources with the Phase I array, it was usually required to use visibility weighting that was closer to uniform weighting \citep[e.g.][]{2017MNRAS.464.1146H}, with the associated loss in sensitivity. This is because the naturally weighted Phase I synthesised beam has a large ``halo'' around the central peak due to the compact core, as can be seen in Fig. \ref{fig:beamcuts}. Conversely, the Phase II extended array has an almost identical beam for both visibility weighting schemes and very low sidelobes away from the phase centre.

Previous deep 154\,MHz images made with the MWA Phase I \citep{2016MNRAS.458.1057O,2016MNRAS.459.3314F} were limited by confusion, with sidelobe confusion contributing to the overall image noise.
The improved angular resolution of the extended array will reduce the classical confusion in continuum images, 
with the precise factor depending on the slope of the differential source count distribution around 1\,mJy \citep{1974ApJ...188..279C}.
Using data from the Phase I MWA and other low frequency telescopes, \citet{2016MNRAS.459.3314F} estimated the classical confusion for Phase I MWA at 154\,MHz to be 1.7\,mJy and the expected reduction in classical confusion to be a factor of 5-10 for Phase II.
The Phase II extended array's improved synthesised beam will also reduce sidelobe confusion. The combination of improved sensitivity, improved classical confusion and reduced sidelobe confusion means an order of magnitude improvement in the noise floor of continuum images from the extended array is expected.

\begin{figure*}
\includegraphics[width=9.2cm]{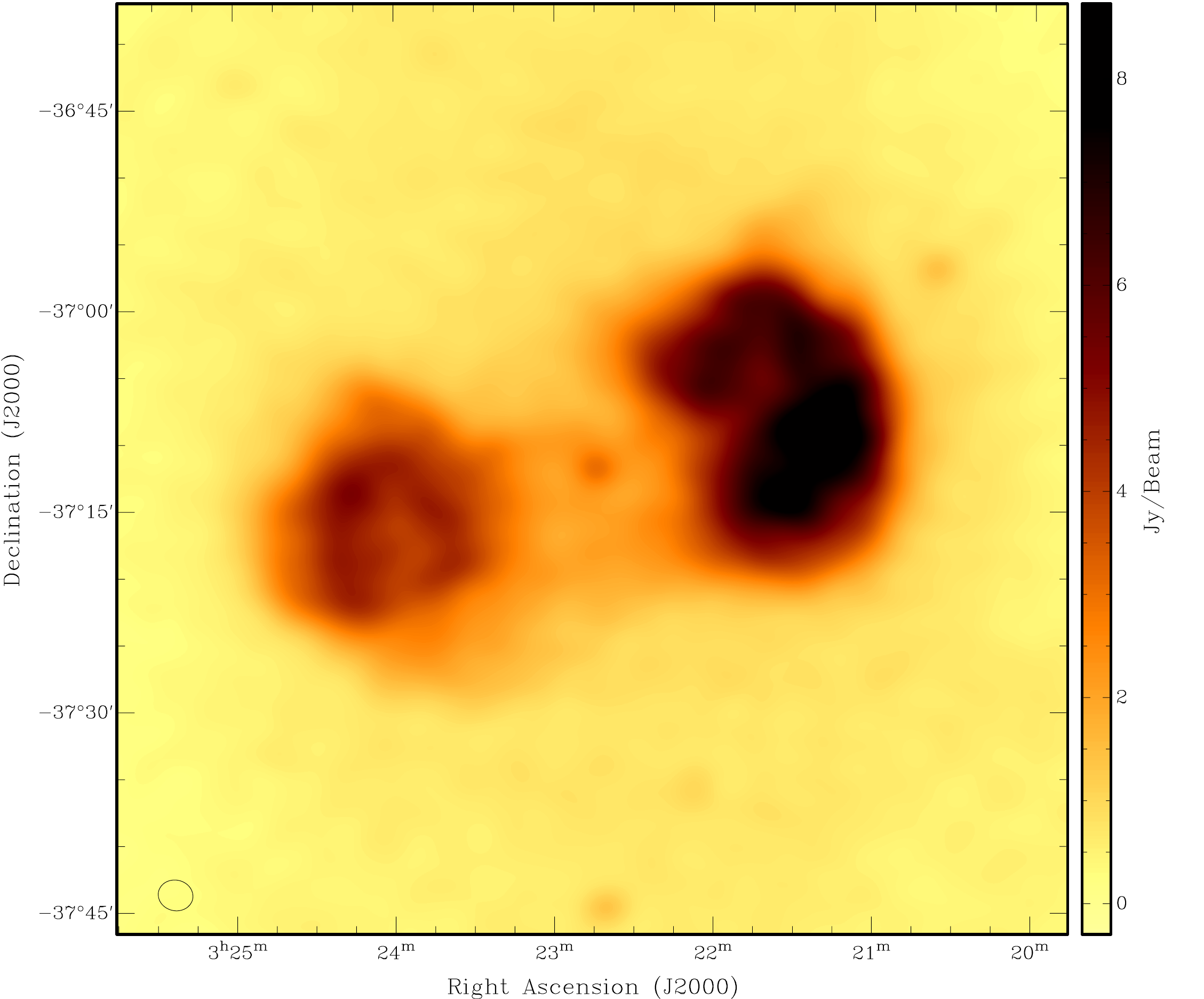}
\includegraphics[width=9.2cm]{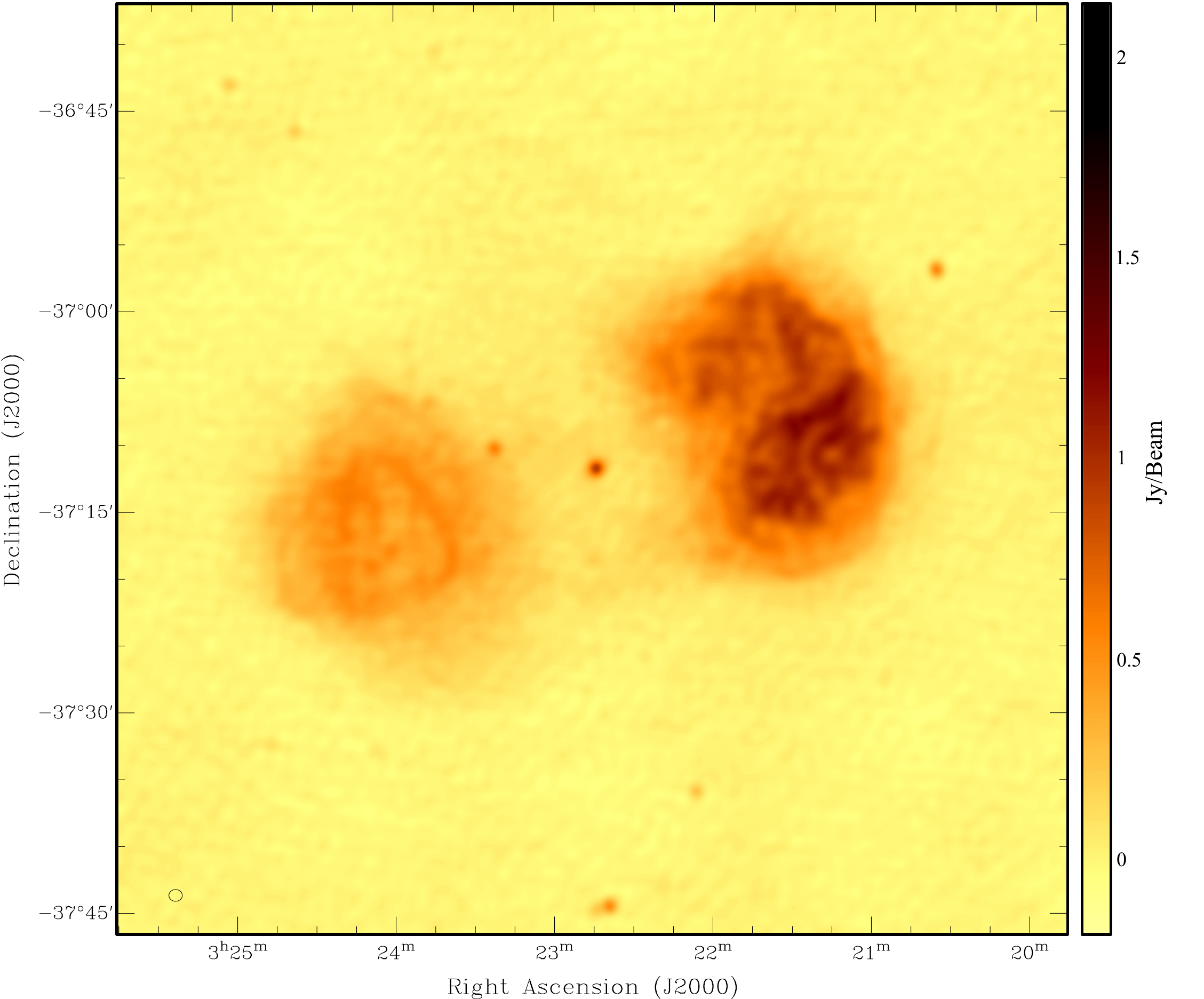}
\caption{185 MHz images of Fornax A taken with the Phase I (left) and Phase II (right) MWA. The ellipse in the lower left of each image shows the synthesised beam sizes, which are $2.6\times2.3$ and $1.0\times0.89$ arcminutes.}
\label{fig:FornaxA}
\end{figure*}

An example of the improvement in resolution achieved between the Phase I and Phase II MWA is shown in Figure \ref{fig:FornaxA} which shows the 185\,MHz Phase~I and Phase~II images of Fornax A.
These images were made with approximately 6 minutes of data for each, with the same frequency, pointing and Local Sidereal Time for each dataset. The same procedure to calibrate and image was used for both datasets using robust 0 visibility weighting.

%\mjh{Need to get some further imaging details from Ben here.}

By design, the compact configuration has very many identical baselines, which are favourable for the EoR power spectrum experiment, but are sub-optimal for imaging applications. The synthesised beam contains regularly spaced grating-like features separated by approximately 9 degrees at 154\,MHz, which are due to the 14\,m separation between tiles in the hexagons. 
Using uniform-like weighting can suppress these grating sidelobes at the expense of approximately 50\% loss in theoretical sensitivity.
Imaging with the compact array will require careful selection of the visibility weighting scheme depending on the application.

\subsection{The link to SKA Low}
It is worth noting that the tradeoffs associated with sensitivity and synthesised beam quality of the MWA (especially the Phase I MWA) are directly relevant to SKA Low. Like Phase I MWA, the proposed layout of SKA1 Low \citep{SKA-TEL-SKO-0000422} contains a densely packed core of approximately 45\% of the antennas with the remaining antennas spread out for angular resolution and $u,v$ coverage. The naturally weighted SKA Low synthesised beam will thus also have a significant halo around the central peak, making deconvolution difficult. High resolution continuum imaging with SKA1 Low will thus likely require significant down-weighting of the shortest baselines with associated reduction in sensitivity.

A future upgrade to the MWA's correlator will allow all 256 antennas to be processed simultaneously. This upgrade will make the MWA's signal processing requirements comparable to SKA Low.
As a precursor to SKA Low, the MWA is an ideal system for proving calibration and imaging pipelines. The ratio of the maximum baseline to station size (hence number of pixels in images) is almost identical, as is the fraction of antennas in the core region.
Expertise and lessons from MWA data processing are thus very relevant for SKA Low.

%%%%%%%%%%%%%%%%%%%%%%%%%
%%%%%%%%%%%%%%%%%%%%%%%%%
\section{CONCLUSION}
\label{sec:conc}

This paper describes the Phase II upgrade of the MWA -- the first major expansion of the radio telescope since it was constructed in 2012. The MWA continues to be a general purpose facility underpinned by its key science programs \citep{2013PASA...30...31B}.
In the first five years of telescope operations, the capabilities of the MWA have been demonstrated across a wide variety of science domains, producing over 100 refereed publications that have generated over 3000 citations.

%\sjt{Why not make this more quantitative and list some of the metrics that we so carefully monitor and are famous for?  Number of papers, citations etc etc.} 

The Phase II upgrade enhances the capabilities of the MWA in angular resolution, brightness sensitivity and sensitivity to the EoR power spectrum while retaining (and even improving) some of the unique features that have made the originally deployed array so successful: the very large field of view combined with the excellent snapshot imaging capability.
The extended configuration, which includes new tiles that double the maximum baseline length, doubles the angular resolution of the array and improves the snapshot $u,v$ coverage. The improved image resolution and synthesised beam will translate to an order of magnitude improvement in continuum image sensitivity due to the reduction in classical and sidelobe confusion.

The compact configuration, which includes the new regularly spaced hexagonal arrays, will improve the sensitivity to the EoR power spectrum by a factor of 3.5. The compact configuration also allows improved and novel array calibration techniques that make use of the hybrid nature of the array layout, which contains redundant and non-redundant baselines.

%\sjt{Perhaps a few words on what may come next for MWA Phase III - more tiles?  Full correlation?  Mention the new correlator and receiver developments?} 
The improvement to the MWA does not stop with the Phase II upgrade; work is ongoing to upgrade several of the digital systems of the MWA. These projects include an upgraded correlator that will correlate all 256 tiles with increased bandwidth, and a new digital receiver that removes the limitations of the existing system. Work is also ongoing to incorporate commensal data processing capability for the Breakthrough Listen project.

Finally, future upgrades of the MWA may include (but are not limited to) retrofitting RFoF transmitters to core tiles to improve the bandpass spectral smoothness; adding more tiles to further extend the maximum baseline; changing the low frequency cutoff of the analogue components to 50\,MHz (since the dipoles have been shown to perform down to 50\,MHz with modified Low-Noise Amplifiers in the Engineering Development Array \citep[`EDA', ][]{2017PASA...34...34W}; and fully integrating the signal paths from the EDA, SKA Low Aperture Array Verification System \citep{8104992} and other SKA Low prototype systems.

\begin{acknowledgements}
The MWA Phase II upgrade project was supported by Australian Research Council LIEF grant LE160100031 and the Dunlap Institute for Astronomy and Astrophysics at the University of Toronto.
Parts of this research were supported by the Australian Research Council Centre of Excellence for All Sky Astrophysics in 3 Dimensions (ASTRO 3D), through project number CE170100013. BM is supported by an ARC Discovery Early-Career Researcher Award, under project number DE160100849.
Parts  of  this  research  were  supported  by  the  Australian  Research 
Council Centre of Excellence for All-sky Astrophysics (CAASTRO), through project number CE110001020.
This scientific work makes use of the Murchison Radio-astronomy Observatory, operated by CSIRO. We acknowledge the Wajarri Yamatji people as the traditional owners of the Observatory site. Support for the operation of the MWA is provided by the Australian Government (NCRIS), under a contract to Curtin University administered by Astronomy Australia Limited. We acknowledge the Pawsey Supercomputing Centre which is supported by the Western Australian and Australian Governments.
\end{acknowledgements}

\bibliographystyle{pasa-mnras}
\bibliography{references}

\end{document}